\documentclass[10pt,conference]{IEEEtran}

\usepackage{latexsym}
\usepackage{cite}
\usepackage{amssymb}
\usepackage{bm}
\usepackage{amsmath, amssymb}
\usepackage[dvips]{graphics}
\usepackage{graphicx}
\usepackage{psfrag}
\usepackage{theorem}
\theoremheaderfont{\bfseries\upshape} \theoremstyle{plain}

\theorembodyfont{\rmfamily}

\theoremheaderfont{\itshape}

\newtheorem{theorem}{Theorem}

\newtheorem{lemma}{Lemma}

\begin{document}
\title{New Bounds for the Capacity Region of the Finite-State Multiple Access Channel
}

\author{
\authorblockN{Haim H. Permuter }
\authorblockA{
Stanford University \\
Stanford, CA \\
 haim1 @ stanford.edu }\and
\authorblockN{Tsachy Weissman}
\authorblockA{
Stanford University \\
Stanford, CA \\
 tsachy @ stanford.edu }\and
\authorblockN{Jun Chen}
\authorblockA{
McMaster University \\
Hamilton, Ontario, Canada\\
 junchen@ece.mcmaster.ca}}

\author{ \authorblockN{Haim Permuter\authorrefmark{1}, Tsachy Weissman\authorrefmark{1}\authorrefmark{2} and Jun Chen\authorrefmark{3},}
\authorblockA{\authorrefmark{1} Department of Electrical
Engineering, Stanford University, Stanford, CA, USA, \{haim1,
tsachy\}@stanford.edu}
\authorblockA{\authorrefmark{2}
Department of Electrical Engineering, Technion, Haifa, Israel}
\authorblockA{\authorrefmark{3}Department of Electrical and Computer Engineering, McMaster
University, Hamilton, Ontario, Canada, junchen@ece.mcmaster.ca}}
\maketitle

\begin{abstract} The  capacity
region of the Finite-State Multiple Access Channel (FS-MAC)  with
feedback that may be an arbitrary time-invariant function of the
channel output samples is considered. We provided a sequence of
inner and outer bounds for this region. These bounds are shown to
coincide, and hence yield the capacity region, of FS-MACs where
the state process is stationary and ergodic and not affected by
the inputs, and for indecomposable FS-MAC when feedback is not
allowed.
 Though the capacity region is `multi-letter' in general, our
results  yield explicit conclusions when applied to specific
scenarios  of interest.
%
%

\end{abstract}


\vspace{-0.1cm}
\section{Introduction}
\vspace{-0.0cm} The Multiple Access Channel (MAC) has received
much attention in the literature. To put our contributions in
context, we begin by briefly describing some of the key results in
the area. The capacity region for the memoryless MAC
 was derived by Ahlswede in \cite{Ahlswede73MAC}.
 Cover and Leung  derived an achievable
region  for a memoryless MAC with feedback in
\cite{CoverL81MACFeedback}. 
Ozarow  derived the capacity of a memoryless Gaussian
MAC  with feedback in \cite{Ozarow84MAC_GaussianFeedback}

In \cite{Kramer98}\cite{Kramer03}, Kramer derived several capacity
results for discrete memoryless networks with feedback. By using the
idea of code-trees instead of code-words, Kramer derived a
`mulit-letter' expression for  the capacity of the discrete
memoryless MAC. One of the main results we develop in the present
paper extends Kramer's capacity result to the case of a stationary
and ergodic Markov Finite-State MAC (FS-MAC), to be formally defined
below.

In \cite{Han98MAC}, Han  used the information-spectrum method in
order to derive the capacity of a general MAC without feedback.
Han also considered the additive mod-$2$ MAC, which we shall use
here to  illustrate the way in which our general results
characterize special cases of interest. In particular, our results
will imply that feedback does not increase the capacity region of
the additive mod-$2$ MAC.

In this work, we consider the  capacity region of the Finite-State
Multiple Access Channel (FS-MAC),  with feedback that may be an
arbitrary time-invariant function of the channel output samples.
We characterize a sequence of inner and outer bounds for this
region and show that it yields the capacity region, for the
important subfamily of FS-MACs.

 Our derivation of the capacity region is rooted in the derivation
 of the capacity of finite-state channels
   in Gallager's book \cite[ch 4,5]{Gallager68}.
More recently,  Lapidoth and Telatar \cite{Lapidoth98Telatar} have
used it in order to derive the capacity of a compound channel
without feedback, where the compound channel consists of a family of
finite-state channels. In particular, they have introduced into
Gallager's proof the idea of concatenating codewords, which we
extend  here to concatenating code-trees.


The paper is organized as follows. We concretely describe the
communication model in Section \ref{sec: Channel Model}. In
Section \ref{sec: directed information}, we introduce the causal
conditioning, directed information and an important idea of
sup/sub-additivity of regions. We state our main capacity results
in Sections \ref{s_FSMAC_feedback} and \ref{s_FSMAC_nofeedback},
and we present a few applications of the capacity results  in
Section \ref{sec:application}. Because of space limitation we do
not provide the proofs. The proofs, with the exceptions of Lemmas
\ref{l_subadditive_region} and \ref{l_subadditive_Rn}, and Theorem
\ref{t_outer_no_feedback}, can be found in the preprint
\cite{Permuter_Weissman_MAC07}.
\vspace{-0.1cm}
\section{Channel Model}
\label{sec: Channel Model} 
\begin{figure}[h]{
 \psfrag{A}[][][0.7]{Encoder 1}
\psfrag{B}[][][0.7]{$x_{1,i}(m_1,z_1^{i-1})$}
\psfrag{C}[][][0.7]{Encoder 2}
\psfrag{D}[][][0.7]{$x_{2,i}(m_2,z_2^{i-1})$}
\psfrag{m1}[][][0.7]{$m_1\;\;$} \psfrag{m2}[][][0.7]{$\;\;\;\;\{
1,...,2^{nR_1}\}\;\;\;\;\;\;\;\;$} \psfrag{m3}[][][0.7]{$m_2
\;\;$} \psfrag{m4}[][][0.7]{$\;\;\;\;\{
1,...,2^{nR_2}\}\;\;\;\;\;\;\;\;$}
 \psfrag{M}[][][0.7]{Finite State
MAC} \psfrag{P}[][][0.7]{$P(y_i,s_i|x_{1,i},x_{2,i},s_{i-1})$}
\psfrag{f}[][][0.7]{$z_{2,i}=f_2(y_{i})$}
\psfrag{f}[][][0.7]{Time} \psfrag{i}[][][0.7]{Invariant}
\psfrag{zf2}[][][0.7]{$z_{2,i}(y_i)$}
\psfrag{zf1}[][][0.7]{$z_{1,i}(y_i)$}
\psfrag{z2}[][][0.7]{$z_{2,i-1}\;$}
\psfrag{z1}[][][0.7]{$z_{1,i-1}\;$} \psfrag{W}[][][0.7]{Decoder}

\psfrag{g}[][][0.7]{Unit} \psfrag{h}[][][0.7]{Delay}
\psfrag{X}[][][0.7]{$\hat m_1(y^n)$} \psfrag{U}[][][0.7]{$\hat
m_2(y^n)$} \psfrag{Y}[][][0.7]{$\hat m_1, \hat m_2$}
\psfrag{D6}[][][0.7]{Function}
\psfrag{v7}[][][0.7]{$z_{i-1}(y_{i-1})$}
 \psfrag{Yi}[][][0.7]{$y_i$}
\psfrag{v6 }[][][0.7]{$\hat m$}\psfrag{w6a}[][][0.7]{Estimated}
\psfrag{w6b\r}[][][0.7]{Message} \centering
\includegraphics[width=9cm]{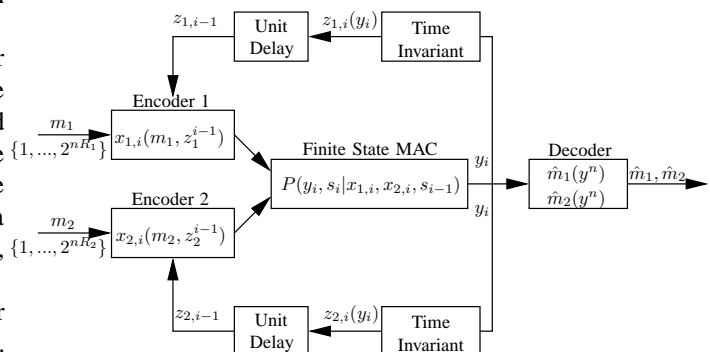}

\centering \caption{Channel with feedback that is a time invariant
deterministic function of the output. \vspace{-0.3cm}
}\label{f_1} }\end{figure}

\vspace{-0.1cm} In this paper, we
consider an FS-MAC (Finite State MAC) with a time invariant
feedback as illustrated in Fig. \ref{f_1}.
The MAC setting consists of two senders and one receiver. Each
sender $l\in\{1,2\}$ chooses an index $m_l$ uniformly from the set
$\{1,...,2^{nR_l}\}$ and independently of the other sender. The
input to the channel from encoder $l$ is denoted by
$\{X_{l1},X_{l2},X_{l3},...\}$, and the output of the channel is
denoted by $\{Y_1,Y_2,Y_3,...\}$. The state at time $i$, i.e.,
$S_i\in \mathcal S$, takes values in a finite set of possible
states. The channel is stationary and is characterized by a
conditional probability $P(y_i,s_i|x_{1i},x_{2i},s_{i-1})$ that
satisfies \vspace{-0.1cm}
\begin{equation}
P(y_i,s_i|x_1^i,x_2^i,s^{i-1},y^{i-1})=P(y_i,s_i|x_{1i},x_{2i},s_{i-1}),
\vspace{-0.1cm}
\end{equation}
where the superscripts denote sequences in the
following way: $x_l^i=(x_{l1},x_{l2},...,x_{li}), \; l\in\{1,2\}$.
We assume a communication with feedback $z_l^i$, where the element
$z_{li}$ is a time-invariant function of the output $y_i$. For
example, $z_{li}$ could equal $y_i$ (perfect feedback), or a
quantized version of $y_i$, or null (no feedback). The encoders
receive the feedback samples with one unit delay.

A code with feedback consists of two encoding functions
$g_l:\{1,...,2^{nR_1}\}\times \mathcal Z_l^{n-1}\to \mathcal
X_l^n,\; l=1,2$,
where the $k{\text th}$ coordinate of $x_l^n\in\mathcal X_l^n $ is
given by the function
\begin{equation}
x_{lk}=g_{lk}(m_l,z_l^{k-1}), \qquad k=1,2,\ldots ,n, \quad l=1,2
\end{equation}
and a decoding function,
\begin{equation}
g:\mathcal Y^n \to \{1,...,2^{nR_1}\} \times \{1,...,2^{nR_2}\}.
\end{equation}
The {\it average probability of error} for
$((2^{nR_1},2^{nR_2},n)$ code is defined as
\begin{equation}
P_e^{(n)}=\frac{1}{2^{n(R_1+R_2)}} \sum_{w_1,w_2}
\Pr\{g(Y^n)\neq(w_1,w_2)|(w_1,w_2) \text{ sent}\}.
\end{equation}
A rate $(R_1,R_2)$ is said to be {\it achievable} for the MAC if
there exists a sequence of $((2^{nR_1},2^{nR_2}),n)$ codes with
$P_e^{(n)}\to 0$. The {\it capacity region} of MAC is the closure
of the set of achievable $(R_1,R_2)$ rates.

%
%
%

\section{Preliminaries}\label{sec: directed information}
\subsection{Causal conditioning and directed information}
Throughout this paper we use the {\it causal conditioning}
notation $(\cdot||\cdot)$. We denote the probability mass function
(pmf) of $Y^N$ causally conditioned on $X^{N-d}$, for some integer
$d\geq0$, as $P(y^N||x^{N-d})$ which is defined as
\begin{equation} \label{e_causal_cond_def}
P(y^N||x^{N-d})\triangleq \prod_{i=1}^{N} P(y_i|y^{i-1},x^{i-d}),
\end{equation}
 (if $i-d\leq0$ then $x^{i-d}$ is set to null). In particular, we
extensively use the cases where $d=0,1$:
\begin{equation}P(y^N||x^{N})\triangleq \prod_{i=1}^{N}
P(y_i|y^{i-1},x^{i})
\end{equation}
\begin{equation}
Q(x^N||y^{N-1})\triangleq \prod_{i=1}^{N} Q(x_i|x^{i-1},y^{i-1}),
\end{equation}
where the letters $Q$ and $P$ are both used for denoting pmfs.

%
The {\it Directed information} was defined by Massey in
\cite{Massey90} as
\begin{equation}
I(X^N \to Y^N )\triangleq \sum_{i=1}^{N} I(X^i;Y_i|Y^{i-1}),
\end{equation}
and in \cite{Kramer98}, Kramer introduced the notation
\begin{equation}
I(X_1^N \to Y^N||X_2^N )\triangleq \sum_{i=1}^{N} I(X_1^i;Y_i|Y^{i-1},X_2^i).
\end{equation}
Directed inofrmation has been widely used in the characterization
of capacity of
 channels
\cite{Kramer98,Tatikonda00,Chen05,Yang05,Permuter06_feedback_submit,Kim07_feedback,ShraderPemuter07ISIT},
and rate distortion function \cite{Pradhan04b,zamir06}. 
Throughout the proofs, we are using several properties of causal
conditioning and directed information. 
We summarize them in the following lemma. 
\begin{lemma}\label{l_hk}
The following four properties
,(\ref{e_property1})-(\ref{e_property4}), hold for any discrete
random vectors $(X_1^N,X_2^N,Y^N)$,
\begin{equation}\label{e_property1}
\hspace{-0.172cm}P(x_1^N,y^N||x_2^N)=P(x_1^N||y^{N-1},x_2^N)P(y^N||x_1^{N},x_2^N).
\end{equation}
\begin{equation}
\hspace{-0.172cm}|I(X_1^N \rightarrow Y^N||X_2^{N})- I(X_1^N
\rightarrow
Y^N||X_2^{N},S)|\leq H(S).
\end{equation}
\begin{eqnarray}
 { \hspace{-0.3cm}\mathcal
I(Q(x_1^N||y^{N-1}),Q(x_2^N||y^{N-1}),P(y^N||x_1^N,x_2^N))\;\;\;\;\;\;\;\;\;\;\;\;}\nonumber
\\ \hspace{0cm} =I(X_1^N\to Y^N||X_2^N),
\end{eqnarray}where $\mathcal I(Q(x_1),Q(x_2),P(y|x_1,x_2))$ denotes the
functional $I(X_1;Y|X_2)$, i.e.,
\begin{eqnarray*}\label{e_calI_def}
&&\hspace{-.8cm}{\mathcal I(Q(x_1),Q(x_2),P(y|x_1,x_2))\triangleq}\\
&&\hspace{-.4cm}
\sum_{y,x_1,x_2}Q(x_1)Q(x_2)P(y|x_1,x_2)\log\frac{P(y|x_1,x_2)}{\sum_{{x}_1^{'}}Q({x}_1^{'})P(y|{x}_1^{'},x_2)}.\;\;\;\;\;\;
\end{eqnarray*}
If there is no feedback, i.e.,
$Q(x_1^N,x_2^N||y^{N-1})=Q(x_1^N)Q(x_2^N)$, then
\begin{equation}\label{e_property4}
I(X_1^N;Y^N|X_2^N) = I(X_1^N \to Y^N||X_2^N).
\end{equation}
\end{lemma}

\subsection{Sup/Sub-additivity, and Convergence of 2D
regions}\label{s_app_supadditive} In this subsection we define
basic operations (summation and multiplication by scalar),
convergence, sup-additivity and sub-additivity of 2D regions.
Furthermore we show that the limit of a sup-additive sequence of
regions converges to the union of all the regions, and the limit
of a sub-additive and convex sequence 2D regions converges to the
intersection of all the regions.

Let $\cal A,B$ be sets in $\mathbb{R}^2$, i.e., $\cal A$ and $\cal
B$ are sets of 2D vectors. The sum of two regions is denoted as
$\cal A+B$ and defined as
\begin{equation}
{\cal A+B}=\{{ {\bf a}+\bf b:\;{\bf a}}\in \cal A, { \bf b}\in
B\},
\end{equation}
and multiplication of a set $A$ with a scalar $c$ is defined as
\begin{equation}
c \mathcal  A=\{c {\bf a}: \;{\bf a}\in \cal A\}.
\end{equation}
A sequence $\{\mathcal A_n\},\; {n=1,2,3,...},$ of 2D regions is
said to {\it converge} to a region $\mathcal A$, written $\mathcal
A=\lim \mathcal A_n$ if
\begin{equation}
\lim\sup \mathcal A_n= \lim\inf \mathcal A_n=\mathcal A,
\end{equation}
where
\begin{eqnarray}\label{e_def_sup_inf_sets}
\lim\inf \mathcal A_n&=&\left\{ {\bf a}:{\bf a}=\lim {\bf a}_n, {\bf a}_n\in \mathcal A_n \right\},\nonumber \\
\lim\sup \mathcal A_n&=&\left\{ {\bf a}:{\bf a}=\lim {\bf a}_k,
{\bf a}_k\in \mathcal A_{n_k} \right\},
\end{eqnarray}
and $n_k$ denotes an arbitrary increasing subsequence of the
integers. 
Let us denote
$\overline{\mathcal A}=\text {cl}\left( \bigcup_{n\geq1} {\mathcal
A}_n \right)$ and $\underline{\mathcal A}= \text
{cl}\left(\bigcap_{n\geq1} {\mathcal A}_n \right).$

We say that a sequence $\{\mathcal A_n\}_{n\geq1}$ is {\it
bounded} if $\sup\{||{\bf a}||: {\bf a}\in \overline {\mathcal
A}\}<\infty$ where $||\cdot||$ denotes a norm in $\mathbb{R}^2$.

\begin{lemma}\label{l_supadditive_region}
Let $ \mathcal A_n$, $n=1,2,...$, be a bounded sequence of sets
in $\mathbb{R}^2$ that includes the origin, i.e., $(0,0)$. If $n
\mathcal A_n$ is sup-additive, i.e., for all $n\geq1$ and all
$N>n$
\begin{equation}\label{e_supadditive_property}
N\mathcal A_N\supseteq n\mathcal A_n+(N-n)\mathcal
A_{N-n}\vspace{-0.1cm}
\end{equation}
 then \vspace{-0.1cm}
\begin{equation}
\lim_{n\to \infty} \mathcal A_n=\overline {\mathcal A}.
\end{equation}
\end{lemma}

\begin{lemma}\label{l_subadditive_region}
Let $\mathcal A_n$, $n=1,2,...$, be a sequence of convex, closed
and bounded sets in $\mathbb{R}^2$. If $n \mathcal A_n$ is
sub-additive, i.e., for all $n\geq1$ and all $N>n$
\begin{equation}\label{e_subadditive_property}
N \mathcal A_N\subseteq n\mathcal A_n+(N-n) \mathcal A_{N-n}
\vspace{-0.1cm}
\end{equation}
 then \vspace{-0.1cm}
 \begin{equation}
 \lim_{n\to \infty}\mathcal A_n=\underline {\mathcal A}.
\end{equation}
\end{lemma}

\begin{figure*}[!t]{
\psfrag{c1}[][][0.8]{$x_1=0$} \psfrag{a1}[][][0.8]{$$}
\psfrag{c2}[][][0.8]{$x_2=1$} \psfrag{a2}[][][0.8]{$i=1$}
\psfrag{c3}[][][0.8]{$x_3=1$} \psfrag{a3}[][][0.8]{$i=2$}
\psfrag{c4}[][][0.8]{$x_4=0$} \psfrag{a4}[][][0.8]{$i=3$}

\psfrag{d1}[][][0.8]{$x_1=0$} \psfrag{a1}[][][0.8]{$$}
\psfrag{d2}[][][0.8]{$x_2=1$} \psfrag{a1}[][][0.8]{$$}
\psfrag{d3}[][][0.8]{$x_2=1$} \psfrag{a1}[][][0.8]{$$}
\psfrag{d4}[][][0.8]{$x_3=0$} \psfrag{a1}[][][0.8]{$$}
\psfrag{d5}[][][0.8]{$x_3=1$} \psfrag{a1}[][][0.8]{$$}
\psfrag{d6}[][][0.8]{$x_3=1$} \psfrag{a1}[][][0.8]{$$}
\psfrag{d7}[][][0.8]{$x_3=1$} \psfrag{a1}[][][0.8]{$$}
\psfrag{da}[][][0.8]{$x_4=0$} \psfrag{db}[][][0.8]{$x_4=1$}
\psfrag{a1}[][][0.8]{$$} \psfrag{a5}[][][0.8]{$i=4$}

\psfrag{k0}[][][0.8]{$\;\;\;\;\;\;z_{i-1}=0$}
\psfrag{k1}[][][0.8]{$\;\;\;\;\;\;z_{i-1}=1$}
\psfrag{k01}[][][0.8]{$\:\:\:\:\:\:\:\:\:\:\:\:$ (no feedback)}

\psfrag{codeword}[][][0.9]{codeword (case of no feedback)}
\psfrag{code-tree}[][][0.9]{code-tree (used in
\cite{Permuter06_feedback_submit})} \psfrag{con
code-tree}[][][0.9]{concatenated code-tree (used here and in
\cite{Permuter_Weissman_MAC07})}

\centerline{\includegraphics[width=17cm]{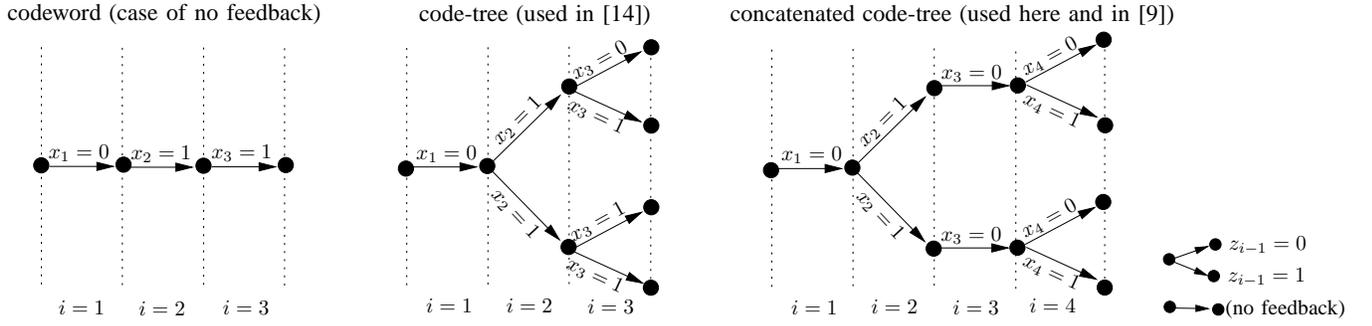}}
\caption{Illustration of coding scheme for setting without
feedback, setting with feedback as used for point-to-point channel
\cite{Permuter06_feedback_submit} and a code-tree that was created
by concatenating smaller code-trees. In the case of no feedback
each message is mapped to a codeword, and in the case of feedback
each message is mapped to a code-tree. The third scheme is a
code-tree of depth 4 created by concatenating two trees of depth
2. } \label{f_codetree} \vspace{-0.40cm}}
\end{figure*}
\section{FS-MAC with time-invariant
feedback}\label{s_FSMAC_feedback}
\subsection{Inner Bound}
 Let $\underline {\mathcal R}_n$ denote the following
region in
$\mathbb{R}_+^2$ (2D set of nonnegative real numbers):
\begin{eqnarray}\label{e_def_underline_Rn}
&&\hspace{-0.7cm}\underline {\mathcal R}_n=\nonumber \\
&&\hspace{-0.7cm}\bigcup_Q
\begin{cases}
R_1 \leq  \min_{s_0} \frac{1}{n}I(X_1^n \to Y^n ||X_2^{n},W,s_0)-\frac{\log |\mathcal S|}{n},\\
R_2 \leq  \min_{s_0} \frac{1}{n}I(X_2^n \to Y^n ||X_1^{n},W,s_0)-\frac{\log |\mathcal S|}{n},\\
R_1+R_2 \leq  \min_{s_0} \frac{1}{n}I((X_1,X_2)^n \to
Y^{n}|W,s_0)-\frac{\log |\mathcal S|}{n},
\end{cases}
\end{eqnarray}
where the union is over the set of all input distributions of the
form $Q(w)Q(x_1^n||z_1^{n-1},w)Q(x_2^n||z_2^{n-1},w)$. Having the
auxiliary random variable $W$ is equivalent to taking the convex
hull of the region. Furthermore, the set of three-inequalities is
equivalent to an intersection of three regions, and $\min_{s_0}$
is equivalent to $\cap_{s_0}$. Hence, an equivalent region is
\begin{eqnarray}\label{e_def_underline_Rn}
&&\hspace{-0.7cm}\underline {\mathcal R}_n=\nonumber \\
&&\hspace{-0.7cm}{\text{conv}} \bigcup_{Q}\bigcap_{s_0}
\begin{cases}
R_1 \leq  \frac{1}{n}I(X_1^n \to Y^n ||X_2^{n},s_0)-\frac{\log |\mathcal S|}{n},\\
R_2 \leq \frac{1}{n}I(X_2^n \to Y^n ||X_1^{n},s_0)-\frac{\log |\mathcal S|}{n},\\
R_1+R_2 \leq  \frac{1}{n}I((X_1,X_2)^n \to Y^{n}|s_0)-\frac{\log
|\mathcal S|}{n},
\end{cases}
\end{eqnarray}
where ${\text{conv}}$ denotes the convex hull, and the input
distribution is of the form
$Q(x_1^n||z_1^{n-1})Q(x_2^n||z_2^{n-1})$. In general, the right
hand side (RHS) of each of the three inequalities that define
${\mathcal R}_n$ can be negative. In such a case, we assume that
the RHS is zero.

%
\begin{theorem}\label{t_inner_bound}
({\it Inner bound.}) For any FS-MAC with time invariant feedback as
shown in Fig. \ref{f_1}, and for any integer $n\geq1$, the region
$\underline {\mathcal R}_n$ is achievable.
\end{theorem}
The proof is similar to the point-to-point FSC with time-invariant
feedback, given in \cite[Sec. V]{Permuter06_feedback_submit}. In
the proof we use Gallager's techniques to analyze the error
probability of a ML decoder of a randomly-generated code. There
are two main differences compared to the point-to-point FSC:
\begin{enumerate}
\item In the case of FSC, only one message is sent, and in the case of FS-MAC, two independent messages are
sent. This requires that we analyze three different types of
errors, and they  
yield three inequalities in the achievable region.
\item For the FS-MAC case, we need to prove the achivebility for a set of input
distributions while for the point-to-point channel it was enough
to prove it  only for the input distribution that achieves the
maximum. Because of this difference, we introduce the idea of
concatenating code-trees (see Fig. \ref{f_codetree}). This
difference influences the encoding scheme and the analysis.

%
%

\end{enumerate}

The following lemma establishes the sub-additivity of
$\{\underline {\cal R}_n\}$.
\begin{lemma}\label{l_supadditive_Rn} ({\it sup-additivity of $\underline {\cal R}_n$.
}) For any FS-MAC, the sequence $\{\underline {\cal R}_n\}$ is
sup-additive. Therefore, $\lim_{n\to\infty} \underline {\mathcal
R}_{n}$ exists, it is an achievable region, and it equals to
$\overline{{\cal R}}$.\vspace{-0.4cm}
\end{lemma}
\subsection{Outer Bound}
The following outer bound is proved
 using Fano's inequality.
\begin{theorem}\label{t_outer_bound}
({\it Outer bound.}) Let $(R_1,R_2)$ be an achievable pair for a
FS-MAC with time invariant feedback, as shown in Fig. \ref{f_1}.
Then, for any $n$ there exists a distribution
$Q(x_1^n||z_1^{n-1})Q(x_2^n||z_2^{n-1})$ such that the following
inequalities hold:
\begin{eqnarray}
R_1 &\leq&  \frac{1}{n}I(X_1^n \to Y^n ||X_2^{n})+\epsilon_n, \nonumber \\
R_2 &\leq&  \frac{1}{n}I(X_2^n \to Y^n ||X_1^{n})+\epsilon_n, \nonumber \\
R_1+R_2 &\leq& \frac{1}{n}I((X_1,X_2)^n \to Y^{n})+\epsilon_n,
\end{eqnarray}
where $\epsilon_n$ goes to zero as $n$ goes to infinity. 
\end{theorem}
This theorem implies that $\liminf {\mathcal R}_n$ is an outer
bound, where ${\mathcal R}_n$ is defined as
\begin{equation}\label{e_def_Rn}
 {\mathcal R}_n=\bigcup_Q
\begin{cases}
R_1 \leq  \frac{1}{n}I(X_1^n \to Y^n ||X_2^{n}),\\
R_1 \leq  \frac{1}{n}I(X_2^n \to Y^n ||X_1^{n}),\\
R_1+R_2 \leq  \frac{1}{n}I((X_1,X_2)^n \to Y^{n}),
\end{cases}
\end{equation}
and the union is over input distributions of the form
$Q(x_1^n||z_1^{n-1})Q(x_2^n||z_2^{n-1})$. \vspace{-0.20cm}
\subsection{Capacity}
Based on the bounds above, we have the following capacity result.
\begin{theorem}\label{t_capacity_feedback} For any FS-MAC
of the form
\begin{equation}\label{e_channel_Markov}
P(y_i,s_i|x_{1,i},x_{2,i},s_{i-1})=P(s_i|s_{i-1})P(y_i|x_{1,i},x_{2,i},s_{i-1}),
\end{equation}
where the state process $S_i$ is stationary and ergodic, the
achievable region is
$\lim_{n\to \infty} \mathcal R_n,$
and the limit exists. \vspace{-0.3cm}
\end{theorem}
\section{FS-MAC without  feedback}\vspace{-0.2cm}\label{s_FSMAC_nofeedback}
The case where there is no feedback is a special case of
deterministic time-invariant feedback in which $z_i$ is null, and
therefore the theorems in the previous section hold for the case
of no feedback. Here we show additional results, which apply only
for the case without feedback. The results include a sequence of
 upper bounds for all FS-MACs, and a capacity
formula for indecomposable FS-MACs.
\subsection{Outer bound}
Let us denote,
\begin{eqnarray}\label{e_def_underline_Rn}
&&\hspace{-0.75cm}\overline {\mathcal R}_n =\nonumber \\
&&\hspace{-0.75cm}\text{conv}\bigcup_{Q}
\begin{cases}
R_1 \leq  \max_P \frac{1}{n}I(X_1^n \to Y^n ||X_2^{n},S_0)+\frac{H(S_0)}{n},\\
R_2 \leq  \max_{P} \frac{1}{n}I(X_2^n \to Y^n ||X_1^{n},S_0)+\frac{H(S_0)}{n},\\
R_1+R_2 \leq  \max_{P} \frac{1}{n}I((X_1,X_2)^n \to
Y^{n}|S_0)+\frac{H(S_0)}{n}\nonumber
\end{cases}
\end{eqnarray}
where the union is over all input distributions of the form
$Q(x_1^n)Q(x_2^n)$, and $\max_P$ denote a maximization over
distribution of the form $P(s_0|x_1^n,x_2^n)$. The sup-additivity
of $\{\overline {\mathcal R}_n\}$ is the key property for
establishing the outer bound.
\begin{lemma}\label{l_subadditive_Rn} ({\it sub-additivity of $\overline {\cal R}_n$.
}) For any FS-MAC, the sequence $\{\overline {\cal R}_n\}$ is
sub-additive, i.e.,
\begin{equation} \label{e_Rn_sub}
(n+l)\overline {\mathcal R}_{n+l}\subseteq n\overline {\mathcal
R}_{n}+l\overline {\mathcal R}_{l}.
\end{equation}
\end{lemma}

\begin{theorem}\label{t_outer_no_feedback} ({\it Outer bound}) For any FS-MAC and all $n\geq1$,  $\overline {\cal
R}_n$ contains the capacity region.\vspace{-0.30cm}
\end{theorem}
\subsection{Capacity}
\begin{theorem}\label{t_no_feedback}
({\it Capacity of FS-MAC without feedback.}) For any indecomposable
FS-MAC without feedback,
\begin{equation}
\lim_{n\to \infty} \underline {\mathcal R}_n=\lim_{n\to \infty}
\overline {\mathcal R}_n,
\end{equation}
and therefore its capacity region is
$\lim_{n\to \infty} \mathcal R_n,$
and the limit exists.\vspace{-0.50cm}
\end{theorem}
\section{Applications}\label{sec:application}\vspace{-0.30cm}
In this section we use the capacity results in order to derive the
following conclusions:
\begin{enumerate}
\item For a stationary ergodic Markovian channels, the
capacity region is zero, if and only if the capacity region with
feedback is zero.

\item For the additive mod-$|\mathcal X|$ MAC, where the noise may have memory:
\begin{enumerate}
\item feedback does not enlarge the capacity;
\item source-channel coding separation holds for lossless
reconstruction.
\end{enumerate}
%
\end{enumerate}

\subsection{Zero capacity}
The first conclusion is given in Theorem \ref{t_capacity_zero}.
\begin{theorem}  \label{t_capacity_zero} For the channel described in
(\ref{e_channel_Markov}), where the state process $S_i$ is
stationary and ergodic, if the capacity without feedback is zero,
then it is also zero in the case that there is feedback.
\end{theorem}

The proof of Theorem \ref{t_capacity_zero} is  based on the fact
that for any MAC
\begin{eqnarray}\label{eqn:0iff}
 \max_{Q(x_1^n||y^{n-1})Q(x_2^n||y^{n-1})} I(X_1^n,X_2^n\to
Y^n)=0
\end{eqnarray}
if and only if
\begin{eqnarray}
\max_{Q(x_1^n)Q(x_2^n)} I(X_1^n,X_2^n\to Y^n)=0,
\end{eqnarray}
and on the fact that for the family of channels that is mentioned
in the theorem, the sequence $\mathcal R_n$ is sup-additive.

For the case of additive Gaussian MAC, one can deduce the result
from the fact that feedback can at most double its capacity region
\cite{Thomas87}. Clearly, Theorem \ref{t_capacity_zero} also holds
for the case of a stationary and ergodic FS-Markov point-to-point
channel because a MAC is an extension of a point-to-point channel.
However, it does not hold for the case of a broadcast channel.

\subsection{Additive mod-$|\mathcal X|$ MAC}
\begin{figure}[h]{
 \psfrag{gamma}[][][0.8]{$\gamma$}
\psfrag{W1}[][][0.8]{$W_1$} \psfrag{W2}[][][0.8]{$W_M$}
\psfrag{X}[][][0.8]{$X_{1,n}(W_1)$}
\psfrag{Y}[][][0.8]{$X_{2,n}(W_2)$} \psfrag{Z}[][][0.8]{$Y_{i}$}
\psfrag{N}[][][0.8]{$V_i$} \psfrag{W3}[][][0.8]{$(\hat W_1,\hat
W_2)$} \psfrag{X2}[][][0.8]{$X_{1,i}(W_1,Y^{n-1})$}
\psfrag{Y2}[][][0.8]{$X_{2,i}(W_2,Y^{n-1})$} \psfrag{A}[][][2]{$$}
\centerline{\includegraphics[width=7cm]{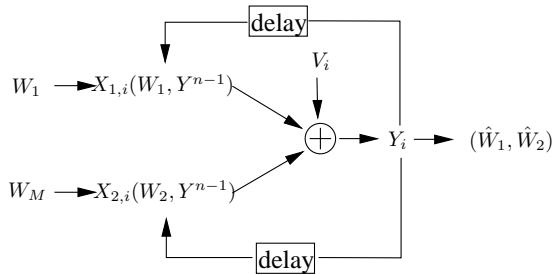}\vspace{-0.20cm}
}


\caption{Additive noise MAC with feedback. The random variables
$X_{1,i},X_{2,i},Y_i,V_i, \; i\in\mathbb{Z^+}$, are from a common
alphabet, and they denote the input from sender 1,2, the output
and the noise at time $i$, respectively. The output satisfies
$y_i=x_{1,i}\oplus x_{2,i} \oplus v_{i}$ where $\oplus$ denotes
addition mod-$|\mathcal X|$. The noise $V_i$, possibly with
memory, is independent of the messages
$W_{1},W_{2}$.\label{f_additive} \vspace{-0.60cm} }
}\end{figure}

In this section we consider the additive mod-$|\mathcal X|$ MAC
with and without feedback. The channel is described in Fig.
\ref{f_additive}. In the binary case, the channel is simply
$Y=X_{1,i}\oplus X_{2,i} \oplus V_i$, where $V_i$ is the binary
noise, possibly with memory, and $\oplus$ denotes addition mod-2.

The following theorem is an extension of of Alajaj's result
\cite{Alajaji95} to the additive MAC .
\begin{theorem}\label{c_additive}
Feedback does not enlarge the capacity region of a discrete
additive (mod-$|\mathcal X|$) noise MAC. Moreover, the capacity
region is given by
\begin{equation}\label{e_cap_region_add}
\sum_{m=1}^M R_m\leq \log q - H(\mathcal V),
\end{equation}
where $H(\mathcal V)$ is the entropy rate of the additive noise.
\end{theorem}

The theorem can be shown to hold for a larger family of  MACs. The
family includes all the MACs that can be represented as
multiplexer followed by a point-to-point channel. The main idea of
the proof is that all three inequalities that defines the region
$\mathcal R_n$ are maximized by uniform and i.i.d distribution,
even if feedback is allowed. 

\subsection{Source-channel coding separation}
Cover, El-Gamal and Salehi \cite{CoverGamalSalehi80MAC_Source}
showed that, in general, the source channel coding separation does
not hold for MACs even for a memoryless channel without feedback.
However, for the case where the MAC is a additive mod-$|\mathcal
X|$, and the goal is to reconstruct the sources losslessly, then
it does hold.

\begin{theorem} {(\it Source-channel coding theorem for a additive mod-$|\cal X|$ MAC.})
 Let $(U_1,U_2)_{n\geq1}$ be a finite alphabet, jointly stationary and ergodic  pair of 
 processes, and let the MAC channel be an additive mod-$|\cal X|$ MAC with stationary and ergodic
 noise. Define $P_e^{(n)}\triangleq \Pr( (\hat U_1^n,\hat U_2^n)\neq ( U_1^n, U_2^n)),$ where $\hat U_1^n,\hat U_2^n$ are the reconstructed sources at the decoder.

{(\it direct part.)} There exists
 a sequence of source-channel codes  with $P_e^{(n)}\to 0$, if $H(\mathcal U_1,\mathcal
 U_2)< \log|\mathcal X|-H(\mathcal V)$, where $H(\mathcal U_1,\mathcal
 U_2)$ is the entropy rate of the sources and $H(\mathcal V)$ is the entropy rate of the noise.

{(\it converse part.)} If $H(\mathcal U_1,\mathcal
 U_2)> \log|\mathcal X|-H(\mathcal V)$, then the probability of error is bounded away from zero, independent of the blocklength. 
\end{theorem}
\vspace{-0.1cm} It is interesting to notice that, even though the
source-channel coding separation theorem holds when the
reconstruction of the sources has to be lossless, the theorem does
not hold when distortion is allowed. Such an example was shown by
Nazar and Gastpar \cite{Nazar_Gastpar07}.\vspace{-0.2cm}
\section{Conclusions and Future Directions} \label{sec: conclusions}
In this paper we have shown that directed information and causal
conditioning emerge naturally in characterizing the capacity
region of FS-MACs in the presence of a time-invariant feedback. We
provided a sequence of inner and outer bounds, and for some large
families of channels we characterize the capacity region  in terms
of  a `multi-letter' expression, which is a first step toward
deriving useful concepts in communication. For instance, we use
this characterization to show that for a stationary and ergodic
Markovian channel, the capacity is zero if and only if the
capacity with feedback is zero. Further, we identify
 FS-MACs for which feedback does not enlarge the capacity region
and for which source-channel separation holds.
%

 One
future direction is to use the characterizations developed in this
paper to explicitly compute the capacity regions of classes of
MACs with memory and feedback (other than the additive
mod-$|\mathcal X|$ channel), and to find optimal coding schemes.
\vspace{-0.4cm}
\bibliographystyle{IEEEtran}
\bibliography{mac_isit08}

\end{document}